\documentclass[aps,preprint,letter,nofootinbib]{revtex4-2}
\usepackage[utf8]{inputenc}
\usepackage{textcomp}
\usepackage{gensymb}
\usepackage{amsmath}
\usepackage{amssymb}
\usepackage{xspace}
\usepackage{nicefrac}
\usepackage{xcolor}
\usepackage{graphicx}
\usepackage{soul}
\usepackage{hyperref}
\usepackage[caption=false]{subfig}
\usepackage{multirow}
\usepackage{cleveref}
\usepackage[titletoc,toc,page]{appendix}
\usepackage{xcolor}

\DeclareMathAlphabet\mathrsfso      {U}{rsfso}{m}{n}

\usepackage{esint}

\begin{document}
\title{Revisiting the implications of Liouville's theorem to the anisotropy of cosmic rays}

\date{\today}

\author{Cain\~a de Oliveira}
\email{caina.oliveira@usp.br}
\author{Leonardo Paulo Maia}
\email{lpmaia@ifsc.usp.br}
\author{Vitor de Souza}
\email{vitor@ifsc.usp.br}
\affiliation{Instituto de F\'isica de S\~ao Carlos, Universidade de S\~ao Paulo, Av. Trabalhador S\~ao-carlense 400, S\~ao Carlos, Brasil.}

\begin{abstract}
We present a solution to  Liouville's equation for an ensemble of charged particles propagating in magnetic fields. The solution is presented using an expansion in spherical harmonics of the phase space density, allowing a direct interpretation of the distribution of arrival directions of cosmic rays. The results are found for chosen conditions of variability and source distributions. We show there are two conditions for an initially isotropic flux of particles to remain isotropic while traveling through a magnetic field: isotropy and homogeneity of the sources. In case isotropically-distributed sources inject particles continuously in time, a transient magnetic induced dipole will appear. This dipole will vanish if the system reaches a steady-state. The formalism is used to analyze the data measured by the Pierre Auger Observatory, contributing to the understanding of the dependence of the dipole amplitude with energy and predicting the energy in which the quadrupole signal should be measured.
\end{abstract}

\maketitle

\section{Introduction}

Liouville's theorem has been applied to study the distribution of sources in the Universe based on the arrival direction of charged particles at Earth. On the way from source to the detector, magnetic fields divert the trajectory of the charged particles. Therefore when detected, they do not point back to their sources. Liouville's theorem connects the measured distribution of arrival direction of particles with the distribution of sources in the Universe by stating that the volume occupied by an ensemble of particles in the phase space will remain constant along the trajectory induced by a Hamiltonian~\citep{arnoldmathematical,goldstein2002classical}. Despite the strong constraint set up by Liouville's theorem, the details on how the distribution of sources rules the distribution of arrival direction of particles for each specific magnetic field is a long subject of debate.

In 1933, \citet{PhysRev.43.87} were the first to use Liouville's theorem to study the effect of Earth's magnetic field in the arrival direction of electrons (Compton effect). The way they used Liouville's theorem in the paper was criticised~\citep{PhysRev.45.835} and corrected~\citep{swann_1933}. Since then, the same formalism has been used to study the effect of the galactic and extragalactic magnetic fields on the arrival directions of charged particles in a wide range of energies~\citep{PhysRev.47.434, Harari_1999, Alvarez-Muniz_2002, Yoshiguchi_2003, Harari_2010, rouille_2014, Eichmann_2020, PhysRevLett.112.021101, Battaner_2015,1266,Aab_2018, Aab_2020_raAnisotropies, deAlmeida:20212Z, AHLERS2017184}. Based on these studies, it has been said that Liouville's theorem guarantees that ``an anisotropy cannot arise through deflections of an originally isotropic flux by a magnetic field"~\citep{1266}. Recently, some authors used detailed simulations of the propagation of charged particles in the Sun's, galactic, and extragalactic magnetic fields to argue that this understanding is not verified for certain configurations of magnetic fields and source distributions~\citep{Lopez-Barquero_2016, Lopez-Barquero_2017,deOliveira_2022_extragalactic}.

We present here a comprehensive derivation of the implications of Liouville's theorem to the arrival directions of particles propagating in magnetic fields, which solves the previous controversies raised in the literature. Using Liouville's theorem, we present in section~\ref{sec:harmonic_exp} a complete solution of the angular power spectrum of arrival directions of particles measured by an observer after propagation through magnetic fields taking into account the distribution of the sources. We show how the magnetic fields can change the direction of an initially generated anisotropic angular distribution (e.g. dipole) and how inhomogeneities in the distribution of sources can generate an anisotropic signal from an initially generated isotropic angular distribution. In section~\ref{sec:monopole_dipole} the evolution of the monopole and the dipole under the effect of a magnetic field is explored. Along the paper, the source setup will change according to the hypothesis we want to test. The source setup gives the initial condition to different situations which will be studied. An ordered magnetic field is assumed to exist through all the space. We show that a spatially homogeneous and isotropic monopole cannot evolve into a dipole in any magnetic field. In section~\ref{sec:quadrupole} we study the evolution of the quadrupole and show the condition in which it could be detectable for a given magnetic field and source distribution.

In section~\ref{sec:monopole_dipole} and~\ref{sec:quadrupole} we use the formalism developed here to explore the arrival direction of ultra-high energy cosmic rays (UHECR) measured by the Pierre Auger Observatory. The source of UHECR is one of the most important open issues in astrophysics today. Recently, anisotropic arrival direction signals have been detected~\citep{1266,Aab_2018,Aab_2020_raAnisotropies,deAlmeida:20212Z}, in particular, a dipolar signal with amplitude $7.3\%$ and direction $\sim115^\circ$ away from the direction of the galactic centre has been measured for particles with energy above $8$~EeV~\citep{deAlmeida:20212Z}. Using the formalism we develop here, we reinterpret the measured UHECR arrival direction dipole, explain the evolution of the measured amplitude with energy, and predict the measurement in the next few years of the still undetected quadrupolar signal.

%----------------------------
%----------------------------
%----------------------------

\section{Arrival direction of charged particles according to Liouville's theorem}
\label{sec:harmonic_exp}

According to Liouville's theorem, the phase flow of Hamiltonian systems preserves volume and for a closed system (constant number of states), the phase space density $f(\mathbf{x},\mathbf{p}, t)$ will also remain constant~\citep{goldstein2002classical,swann_1933,AHLERS2017184},
\begin{equation} \label{eq:liouville}
    \frac{d f(\mathbf{x},\mathbf{p}, t)}{dt} = \partial_t f + \frac{d{\mathbf{x}}}{dt} \cdot \mathbf{\nabla} f + \frac {d{\mathbf{p}}}{dt} \cdot \mathbf{\nabla}_p f = 0,
\end{equation}
where $\mathbf{\nabla} = (\partial_x, \partial_y, \partial_z)$, $\mathbf{\nabla}_p = (\partial_{p_x}, \partial_{p_y}, \partial_{p_z})$, $\mathbf{x}$ the position, and $\mathbf{p}$ the momentum. Relativistic particles with energy $E \approx |\mathbf{p}| c$ and charge $q$ propagating only under the effect of magnetic field $\mathbf{B}$ have force $\frac{d{\mathbf{p}}}{dt}=\frac{q}{|\mathbf{p}|c} \mathbf{p} \times \mathbf{B}$.

If the number of particles in the system increases, equation~\ref{eq:liouville} is modified to include a source term $\Xi(\mathbf{x},\mathbf{p}, t)$ that injects particles of momentum $\mathbf{p}$ in a position $\mathbf{x}$ and time $t$,
\begin{equation} \label{eq:liouville+source}
    \frac{d f(\mathbf{x},\mathbf{p}, t)}{dt} = \partial_t f + \frac{d{\mathbf{x}}}{dt} \cdot \mathbf{\nabla} f + \frac {d{\mathbf{p}}}{dt} \cdot \mathbf{\nabla}_p f = \Xi(\mathbf{x},\mathbf{p}, t),
\end{equation}

In order to solve the evolution of the phase space density $f$ according to the Liouville equation and to establish the connection with the arrival directions of the particles in space, we write $f$ and $\Xi$ in the spherical harmonics basis,
\begin{equation}
  \label{harmonic_exp}
  f(\mathbf{x},\mathbf{p}, t) = \sum_{\ell,m} a_{\ell m} (\mathbf{x},|\mathbf{p}|, t) Y_{\ell m}(\theta_p,\phi_p),
\end{equation}
\begin{equation}
  \label{harmonic_exp_S}
  \Xi(\mathbf{x},\mathbf{p}, t) = \sum_{\ell,m} S_{\ell m} (\mathbf{x},|\mathbf{p}|, t) Y_{\ell m}(\theta_p,\phi_p).
\end{equation}
Using the definition of force above, the multipolar expansion of $f$ and the angular momentum operator in the velocity subspace ( $\mathbf{L}=-i \mathbf{p} \times \mathbf{\nabla}_p$), equation~\ref{eq:liouville+source} can be rewritten
\begin{equation}
  \label{eq:liouville-projected}
  \sum_{\ell,m} \Bigg[ (\partial_t a_{\ell m}) Y_{\ell m} + c  (\hat{\mathbf{p}} \cdot \mathbf{\nabla} a_{\ell m}) Y_{\ell m} - i\frac{qc}{E} a_{\ell m} \mathbf{B} \cdot \mathbf{L} Y_{\ell m} \Bigg] = \sum_{\ell,m} S_{\ell m} Y_{\ell m},
\end{equation}
where $\mathbf{v} \approx c \hat{\mathbf{p}}$.

Note that $f(\mathbf{x},\mathbf{p},t)$ is a density in phase space. Its integral in the momentum subspace returns the particle density in the physical space. All the multipoles $a_{\ell m}$ derived from $f$ as above constitute a density field in space. To evaluate the multipole measured by an observer, it is necessary to integrate its respectively density field $a_{\ell m}$ in the observer region.

The components ($a_{\ell m}$) give the evolution of each multipole and correspond to the coefficients of the spherical harmonics analysis of the arrival directions of particles in space. The evolution of each component is obtained by projecting onto the corresponding element of the dual basis ($Y^*_{\ell' m'})$:
\begin{equation}
  \label{eq:liouville_expanded}
  \partial_t a_{\ell m} = - c \sum_{\ell' m'}T_{\ell m}^{\ell' m'} a_{\ell' m'} - \frac{qc}{E} \sum_{m'} \mathrsfso{B}_{m}^{m'} a_{\ell m'} + S_{\ell m}
\end{equation}
the details are given in appendix \ref{app:multipoles} where we have defined the transition tensor $T_{\ell m}^{\ell' m'}$ and the magnetic tensor $\mathrsfso{B}_{m}^{m'}$:
\begin{multline*}
  T_{\ell m}^{\ell' m'} := \sqrt{\frac{2\ell'+1}{2\ell + 1}} C^{1 0| \ell' 0}_{\ell 0} \Bigg[\frac{ C^{1\ -1 | \ell' m'}_{\ell m}\partial_+ -C^{1 1 | \ell' m'}_{\ell m}\partial_-}{\sqrt{2}} + C^{1 0 | \ell' m'}_{\ell m}\partial_z  \Bigg]   \; \; \mathrm{and} \\
 \mathrsfso{B}_{m}^{m'} := -i \bigg[ \frac{B_+}{2} \sqrt{\ell(\ell+1) - m(m+1)} \delta_{m+1}^{m'}
    +\frac{B_-}{2} \sqrt{\ell(\ell+1) - m(m-1)} \delta_{m-1}^{m'}
    + m B_z \delta_{m}^{m'} \bigg],
\end{multline*}
with $\partial_\pm := \partial_x \pm i\partial_y$, $B_\pm := B_x \pm iB_y$ and $C_{\ell m}^{j k | \ell' m'} \equiv \langle j k\ \ell' m' | \ell m \rangle$ are the Clebsch-Gordan coefficients.

Equations~\ref{eq:liouville_expanded} give the time evolution of each multipole during the propagation of the particles through a magnetic field. Two general conclusions can be drawn from this equation. Spatial inhomogeneities of one multipole can generate consecutive multipoles (e.g. inhomogeneous monopole generates dipole), corresponding to transitions $\ell \mapsto \ell \pm 1$ coupled by the transition tensor. The magnetic field can cause only spatial rotations of the multipoles (e.g. rotation of the dipole direction), corresponding to transitions $m \mapsto m, m \pm 1$ coupled by the magnetic tensor. This is also shown in Appendix~\ref{app:amplitude:evolution} in an alternative way which includes the calculation of the evolution of the angular power spectrum $C_{\ell}$.

Through this paper, we assume that the sources emit particles isotropically,  $S_{\ell m} = S \delta_{\ell 0}\delta_{m 0}$.

%========================================================================
\section{Monopole and Dipole evolution} 
\label{sec:monopole_dipole}

Equation \ref{eq:liouville_expanded} gives the evolution of the monopole ($\Phi \rightarrow \ell=0$) and dipole ($\mathbf{D} \rightarrow \ell=1$) 
\begin{align}
    &\partial_t \Phi (\mathbf{x},t) = - c \mathbf{\nabla} \cdot \mathbf{D}(\mathbf{x},t) + S(\mathbf{x},E,t)\label{eq:monopole}\\
    &\partial_t \mathbf{D}(\mathbf{x},t) = -\frac{c}{3} (\mathbf{\nabla} \Phi + \mathbf{\nabla} \cdot \stackrel{\leftrightarrow}{Q} ) + \frac{qc}{E} \mathbf{D}(\mathbf{x},t) \times \mathbf{B}(\Vec{x}) \; ,
    \label{eq:dipole_old}
\end{align}
where $\Phi = \sqrt{4\pi} a_{00}$ and $\mathbf{D} = \sqrt{\frac{2\pi}{3}} \Big( (a_{1-1} - a_{11}), -i (a_{1-1} + a_{11}), \sqrt{2} a_{10}  \Big)$, as usual. $\stackrel{\leftrightarrow}{Q}$ denotes the traceless quadrupole tensor.

As discussed above, $\Phi$ and $\mathbf{D}$ are density fields in space. The measured multipole is the integral over the observer region $\mathcal{O}$ of the densities, with interpretations
\begin{align*}
    &\int_\mathcal{O} \Phi (\mathbf{r}) d^3\mathbf{r} = N\text{, the detected number of particles,}\\
    &\int_\mathcal{O} \mathbf{D}(\mathbf{r}) d^3\mathbf{r} = \mathbf{D}_{obs}\text{, the dipole measured by an observer.}
\end{align*}
In all cases discussed through this paper, we assume a spherical observer with radius $R$. As the dipole amplitude scales with the number of particles detected, it is suitable normalize the dipole by $N$, and we will define the observed dipole $\mathcal{D}$ as
\begin{equation*}
    \mathcal{D} = \frac{1}{\int_\mathcal{O} \Phi(\mathbf{r}) d^3 \mathbf{r}} \int_\mathcal{O} \mathbf{D}(\mathbf{r}) d^3 \mathbf{r}.
\end{equation*}

Equations~\ref{eq:monopole} and \ref{eq:dipole_old} show more clearly the hierarchy in the multipoles production given only by consecutive poles ($\ell \pm 1$) as discussed above. Equation~\ref{eq:monopole} shows that the evolution in time of the monopole ($\ell = 0$) depends only on the divergence of the dipole ($\ell = 0 + 1 = 1$). Equation~\ref{eq:dipole_old} shows that the evolution in time of the dipole ($\ell = 1$) depends on the spatial variation of the monopole ($\ell = 1 - 1 = 0$) and on the divergence of the quadrupole ($\ell = 1 + 1 = 2$). Therefore, for instance, a contribution of an initial quadrupole to the time evolution of the monopole is propagated via the generation of a dipole which in turn can influence the time evolution of the monopole. In other words, the time evolution of one pole depends on its first neighbors, and the influence of further poles is reduced.

Following this argument, we consider the case of a dominant initial monopole and that the quadrupole contribution to the dipole can be neglected ($\stackrel{\leftrightarrow}{Q}=0$). Equation \ref{eq:dipole_old} becomes
\begin{align}
    &\partial_t \mathbf{D}(\mathbf{x},t) = -\frac{c}{3} \mathbf{\nabla} \Phi + \frac{qc}{E} \mathbf{D}(\mathbf{x},t) \times \mathbf{B}(\Vec{x}) \; ,
    \label{eq:dipole}
\end{align}

In the following, we solve equations~\ref{eq:monopole} and~\ref{eq:dipole} for four particular cases of source distribution: a)  stationary condition, b) transient solution for continuously emitting sources, c) spatially homogeneous and isotropic monopole, and d) spatially isotropic monopole. In the cases where isotropy is imposed, we consider the observer centered at r=0.

%%%%%
\vspace{0.5cm}
\textit{(a) Stationary condition:}

As can be seen in equation \ref{eq:dipole}, a particular balance between the monopole gradient and the magnetic fields can freeze the dipole in time, generating a stationary condition: $\Phi(\mathbf{x},t) = \Phi(\mathbf{x})$ and $|\mathbf{D}(\mathbf{x},t)| = D (\mathbf{x})$.  This situation is possible regardless of the sources distribution, with the only requirement that the particle  distribution reach an stationary state. The stationary state is expected, after a long time, for impulsive sources distributed in a very extended (formally infinity) region or for continuously-emitting sources. Note that for a generic source distribution the existence of a dipole is, in general, expected.

For sources injecting particles impulsively, imposing stationarity ($\partial_t D(\mathbf{x},t) = 0$) in equation \ref{eq:dipole}, we obtain $\frac{q}{E} B D \sin \theta \sim \frac{1}{3} \frac{\Phi}{L}$, where $L$ is the typical spatial variation scale of the monopole and $\theta$ the angle between the stationary dipole and the magnetic field. From this relation, it is possible to estimate
\begin{equation} 
\label{eq:stat:dip}
\frac{D}{\Phi} \sim \frac{1}{3\sin \theta}\frac{1}{B/\text{nG}}\frac{1}{L/\text{Mpc}}\frac{E/\text{EeV}}{Z} \; ,
\end{equation}
where we have used $q = Ze$. This equation shows that the dipolar amplitude at a particular position in space depends on the inverse of the magnetic field intensity, on the inverse of the spatial variation scale of the original monopole and most importantly grows linearly with energy.

In the presence of continuously emitting sources, a stationary condition also can be reached. Imposing stationarity in equation ~\ref{eq:monopole}, we obtain $\nabla \cdot D = c^{-1} S(\mathbf{x},E)$, whose solution is
\begin{equation} \label{eq:dip_stat_source}
    \mathbf{D} = \frac{1}{4\pi c} \int d^3 r' S(\mathbf{r',E}) \frac{\mathbf{r} - \mathbf{r'}}{|\mathbf{r} - \mathbf{r'}|^3}.
\end{equation}
Taking the divergence of equation \ref{eq:dipole}, we find
\begin{align}
    \nabla^2 \Phi = \frac{3q}{E} \Big( \mathbf{B} \cdot \nabla \times \mathbf{D} - \mathbf{D} \cdot \nabla \times \mathbf{B} \Big),
\end{align}
which can be solved to obtain
\begin{align}
    \Phi = \frac{3q}{4\pi E} \int d^3 \mathbf{r'} \frac{1}{|\mathbf{r} - \mathbf{r'}|}\Big( \mathbf{B} \cdot \nabla \times \mathbf{D} - \mathbf{D} \cdot \nabla \times \mathbf{B} \Big) \; .
    \label{eq:monopole_stat_source}
\end{align}

In the case in which the sources have identical energy spectrum, equations~\ref{eq:dip_stat_source} and \ref{eq:monopole_stat_source} implies that the linear dependence of~$D/\Phi$ with the energy remains valid.

In case that the sources are isotropically distributed, $S(\mathbf{x},E) = S(r,E)$, the dipole is purely radial
\begin{equation}
    \mathbf{D}
    = \frac{\hat{\mathbf{r}}}{rc} \int dr' \frac{(r')^2 S(r')}{r+r'} = D(r) \mathbf{\hat{r}},
\end{equation}
and then no anisotropy will be measured in the stationary condition to isotropically distributed sources, since the observed dipole will be
\begin{equation}
    \mathcal{D} = \frac{1}{N} \int_{\mathcal{O}} D(r) \mathbf{\hat{r}} d^3\mathbf{r} = \frac{1}{N}\int_0^R r^2 D(r) dr \int d\Omega \mathbf{\hat{r}} = 0,
\end{equation}
where we consider a observer centered in the origin. The lack of anisotropy comes from the fact that the angular integral of the radial unit vector is null. This can be interpreted as, after a long time the contribution of particles from all directions becomes equal, eliminating any special arrival direction.

%%%%%
\vspace{0.5cm}
\textit{(b) Transient solution in the presence of sources:}

Consider the case where the sources inject particles continuously from $t=0$ onwards. While $c \nabla \cdot \mathbf{D} \ll S $ the monopole evolution in equation \ref{eq:monopole} can be approximated by $\partial_t \Phi \approx S(\mathbf{x},E)$, and then $\Phi \approx S(\mathbf{x},E) t$. The solution of equation \ref{eq:dipole} can be written as
\begin{equation}
    \mathbf{D}(\mathbf{r},t) = -\frac{c}{3} |\nabla S(\mathbf{r},E)| \Bigg[%
    \frac{t^2}{2} \hat{s}%
    - \Bigg( \frac{t}{\omega} - \frac{\sin \omega t}{\omega^2} \Bigg) \hat{b} \times \hat{s}%
    + \Bigg( \frac{t^2}{2} - \frac{1}{\omega^2} + \frac{\cos \omega t}{\omega^2} \Bigg) \hat{b} \times \hat{b} \times \hat{s}
    \Bigg]
\end{equation}
where $\omega = \frac{qc}{E} B(\mathbf{r})$, $\hat{s} = \nabla S/|\nabla S|$, $\hat{b} = \mathbf{B}(\mathbf{r})/B(\mathbf{r})$. The solution is valid for a time interval such that $c \nabla \cdot D \ll S$, or $t \ll \sqrt{2\mu_s^{-1}} L_s/c \sim 7 (L_s/\text{Mpc})$~Myr, where $L_s$ is the typical scale of variation of $S$ and $\mu_s = \mathbf{\hat{b}} \cdot \mathbf{\hat{s}}$. In this situation, a non-zero dipole vector can be measured in an observer sphere of radius $R$, even if the source distribution is isotropic, $S=S(r)$.
The magnetic field breaks the isotropy on the trajectory of the particles, which will take different times to reach the observer coming from different directions. This generates a higher flux of particles coming from special directions.

%%%%%
\vspace{0.5cm}
\textit{(c) Spatially homogeneous and isotropic source distribution:}
Consider the case where continuously-, homogeneously-, and isotropically-distributed identical sources inject particles isotropically at $t=0$. This initial condition corresponds to a phase space density $f$ given only by a monopole identical in all points of the space. The solution for $\Phi$ is a constant in space $\mathbf{\nabla}\Phi = 0$. 

From equation~\ref{eq:dipole}, the dipole evolution is given by $\partial_t \mathbf{D} = \frac{qc}{E} \mathbf{D}\times \mathbf{B}$ which leads to $\frac{1}{2} \partial_t D^2 = \frac{ec}{E} \mathbf{D} \cdot (\mathbf{D}\times \mathbf{B}) =0$ and therefore $\partial_t D = 0$ independently of any assumption on the magnetic field. If the dipole remains constant in time, higher orders poles cannot be generated given the argument of hierarchy discussed above. In summary, a flux of particles emitted isotropically by sources distributed continuously, homogeneously and isotropicaly will be detected isotropically by an observer in any point of the space and at any time.

%%%%%
\vspace{0.5cm}
\textit{(d) Spatially isotropic source distribution:}

Consider the case where isotropically distributed identical sources inject particles isotropically impulsively at $t=0$.  This initial condition corresponds to a phase space density $f$ given only by a monopole which can vary in space, $\Phi$ must be isotropic but not necessarily homogeneous, i.e. $\Phi = \Phi (r)$. 

It is our intention to show that under this condition (isotropic sources), an observer can detected an anisotropic arrival direction of particles for some configurations of magnetic fields. We could not find general analytical solutions for equations~\ref{eq:monopole} and~\ref{eq:dipole} for any magnetic field configuration. We will solve these equations for a particular case of magnetic field intensity known as weak field solutions.

We follow the usual perturbative method writing $\Phi = \Phi^{(0)} + \epsilon \Phi^{(1)} + \mathcal{O} (\epsilon^2)$, and $\mathbf{D} = \mathbf{D}^{(0)} + \epsilon \mathbf{D}^{(1)} + \mathcal{O} (\epsilon^2)$, $qc\mathbf{B}/E \mapsto \epsilon qc\mathbf{B}/E$, and collecting terms of the same order in $\epsilon$, equations~\ref{eq:monopole} and~\ref{eq:dipole} become
\begin{align}
    &\partial_t \Phi^{(0)} = - c \nabla \cdot \mathbf{D}^{(0)}
    &&\partial_t \mathbf{D}^{(0)} = - \frac{c}{3} \mathbf{\nabla} \Phi^{(0)} \label{eq:order_zero}
    \\
    &\partial_t \Phi^{(1)} = - c \nabla \cdot \mathbf{D}^{(1)}
    &&\partial_t \mathbf{D}^{(1)} = - \frac{c}{3} \mathbf{\nabla} \Phi^{(1)} - \frac{qc}{E} \mathbf{B} \times \mathbf{D}^{(0)} \label{eq:order_one}.
\end{align}
As shown in Appendix \ref{app:weak-field}, given the initial condition of a monopole with spherical symmetry ($\Phi = \Phi (r)$), these equations can be solved
\begin{align}
    &\Phi^{(0)} (\mathbf{r},t) = \frac{F(k r+\omega t)}{r},\\
    &\mathbf{D}^{(0)} (\mathbf{r},t) = \frac{g(r,t)}{r^2} \hat{\mathbf{r}},\\
    &\Phi^{(1)} (\mathbf{r},t) = \Phi^{(1)}(r,t),\\
    &\mathbf{D}^{(1)} (\mathbf{r},t) = -\frac{c}{3} \int_0^t \partial_r \Phi^{(1)}(r,\tau) d\tau \; \hat{\mathbf{r}} + \frac{qc}{E} \frac{\hat{\mathbf{r}} \times \mathbf{B}(\mathbf{r})}{r^2} \int_0^t g(r,\tau) d\tau,
\end{align}
where $\omega^2 = \frac{1}{3} c^2 k^2$, $F(kr + \omega t)$ is determined by the initial conditions, and $g(r,t)$ is a function of $F(kr+\omega t)$.

An observer with a given radius $R$ will measure a dipolar signal given by all particles coming from all directions
\begin{align}
    \mathbf{\mathcal{D}}^{(0)} (r,t) &= \frac{\int_\mathcal{O} \mathbf{D}^{(0)} d^3 \mathbf{r}}{\int_\mathcal{O} \Phi^{(0)} d^3 \mathbf{r}}  = \frac{\int_0^R  g(r,t) dr}{\int_\mathcal{O} \Phi^{(0)} d^3 \mathbf{r}} \oiint \hat{\mathbf{r}} d\Omega = 0,
\end{align}
and the first-order perturbation to the measured dipole is given by
\begin{align*}
    %\mathbf{\mathcal{D}}^{(1)} (r,t) &= \frac{\oiint \mathbf{D}^{(1)} dS}{\oiint \Phi^{(0)} dS} = -\frac{c}{3} \frac{\int_0^t \partial_r \Phi^{(1)}(r,\tau) d\tau}{4\pi r^2 \Phi^{(0)}} \oiint \hat{\mathbf{r}} dS + \frac{qc}{E} \int_0^t g(r,\tau) d\tau \frac{\oiint \hat{\mathbf{r}} \times \mathbf{B}(\mathbf{r}) dS}{4\pi r^4 \Phi^{(0)}}\\
    %&= \frac{qc}{E} \frac{1}{4\pi r^3 F(kr+ \omega t)} \Bigg[ \int_0^r s^2ds \int_{-1}^1 d(\cos \theta) \int_0^{2\pi} d\phi  \nabla \times \mathbf{B}(\mathbf{s}) \Bigg] \int_0^t g(r,\tau) d\tau.
    \mathbf{\mathcal{D}}^{(1)}& (r,t) = \frac{\int_\mathcal{O} \mathbf{D}^{(1)} d^3 \mathbf{r}}{\int_\mathcal{O} \Phi^{(0)} d^3 \mathbf{r}}\\%
    &= -\frac{c}{3} \frac{\int_0^R \int_0^t \partial_r \Phi^{(1)}(r,\tau) d\tau r^2dr}{\int_\mathcal{O} \Phi^{(0)} d^3 \mathbf{r}} \int_{\Omega} \hat{\mathbf{r}} d\Omega + \frac{qc/E}{\int_\mathcal{O} \Phi^{(0)} d^3 \mathbf{r}} \int_0^R \int_0^t \frac{g(r,\tau)}{r^2} d\tau \int_{\Omega} \hat{\mathbf{r}} \times \mathbf{B}(\mathbf{r}) r^2 d\Omega dr\\
    &= \frac{qc}{E} \frac{1}{4\pi \int_0^R r^2 \Phi^{(0)}dr} \int_0^R dr\Bigg[ \int_0^r s^2ds \int_{-1}^1 d(\cos \theta) \int_0^{2\pi} d\phi  \nabla \times \mathbf{B}(\mathbf{s}) \Bigg] \int_0^t \frac{g(r,\tau)}{r^2} d\tau.
\end{align*}
Using the definition $\langle \mathbf{\nabla} \times \mathbf{B} \rangle (r) = \big( \frac{4}{3}\pi r^3 \big)^{-1} \int (\mathbf{\nabla} \times \mathbf{B}) d^3\mathbf{s}$,
\begin{equation}
\label{eq:dipole:weak:field}
    \mathbf{\mathcal{D}}^{(1)} (r,t) = \frac{qc}{3 E} \frac{\int_0^R \langle \mathbf{\nabla} \times \mathbf{B} \rangle \int_0^t r g(r,\tau) d\tau dr}{\int_0^R r^2 \Phi^{(0)}dr},
\end{equation}
and the dipole measure by an observer ignoring higher order in $\epsilon$ is given by $\mathcal{D} = \epsilon \mathcal{D}^{(1)} \neq 0$. This equation shows that an isotropic and inhomogeneous source distribution can generate a dipolar signal in an observer. This conclusion is valid for weak field approximation and radially distributed sources. No assumption about the structure of the magnetic field was considered in the calculations. The dipolar signal is generated by the interactions of the gradient of the initial inhomogeneous monopole with the magnetic field.

%%%%%
\subsection{Ultra-high energy cosmic rays and Liouville's theorem}

In this section, we discuss the solutions of the Liouville equation presented above in the context of the arrival directions of UHECR. The arrival directions of UHECR on Earth as measured by the Pierre Auger Observatory are dominated by a monopole. A dipole with amplitude $7 \pm 1$\% and direction ($95^\circ \pm 8^\circ,-36^\circ \pm 9^\circ$) in equatorial coordinates is also detected, in the energy range above $8$~EeV. The quadrupole and poles of higher degrees are negligible~\citep{deAlmeida:20212Z}. The dependence of the dipole with energy was measured as $D \propto E^{0.98 \pm 0.15}$ which agrees very well with the linear dependence calculated above and summarized in equation~\ref{eq:stat:dip}. 

Using equation~\ref{eq:dipole} and imposing $\mathbf{B}\times \mathbf{\nabla} \Phi \neq 0$, it is also possible to understand how the dipole direction may not point directly to the source but will suffer precession in the extragalactic magnetic field with an angular frequency $qBc/E$, or a period $(20/Z)(E/\text{EeV}) (\text{nG}/B)$~Myr.

\begin{figure}
  \centering
  \includegraphics[width=0.5\columnwidth]{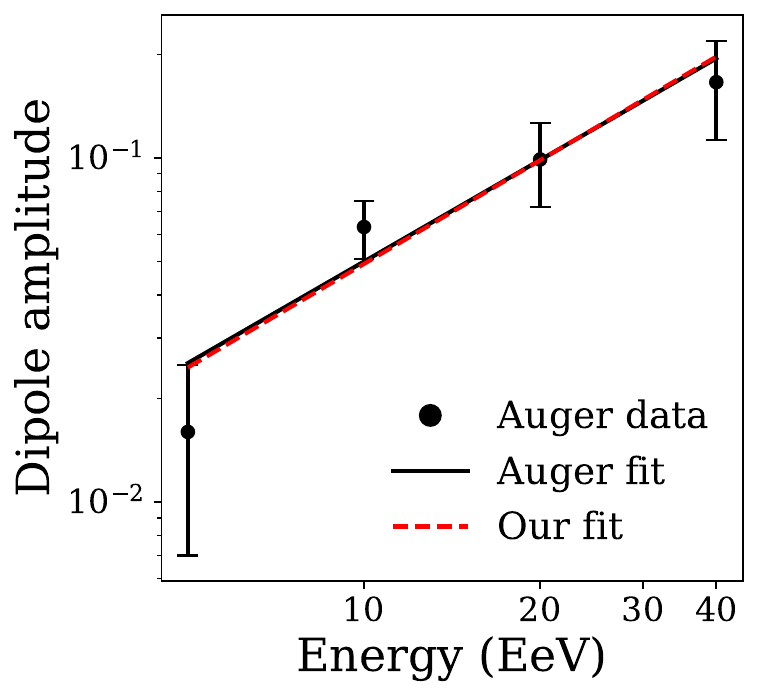}
  %161 x 297 mm = 297 words
  \caption{Dipole amplitude as a function of energy published by the Pierre Auger Collaboration~\citep{deAlmeida:20212Z}. The data were fitted by the Pierre Auger Collaboration~\citep{deAlmeida:20212Z} as shown by the black line. We fit the data using our model as shown by the red dashed line.}
  \label{fig:dipole}
\end{figure}

Figure~\ref{fig:dipole} shows a fit of equation~\ref{eq:stat:dip} to the Pierre Auger published data. Considering protons ($Z=1$), the fit gives $BL\sin \theta \sim (68 \pm 9)$~nG~Mpc. Given independent measurements on the extragalactic magnetic field intensity $B \sim 0.2-22$~nG~\citep{10.1093/mnras/stab3495, Bray_2018, PhysRevLett.116.191302} we find $L \sim 3-350$~Mpc. Considering iron nuclei ($Z = 26$), the fit gives $L \sim 0.1 - 15$~Mpc. From its definition (equation~\ref{eq:stat:dip}), $L$ is the typical spatial variation scale of the monopole, therefore it correlates with the average distance of the sources. Estimations of the sources density~\citep{Aab_2018} lead to the characteristic distance of the source $\sim 10 - 50$~Mpc. However, it is important to note that the dipole amplitude is a function of the rigidity ($E/Z$), and considering that the UHECR composition might vary with energy~\citep{Aab_2018}, a departure from a linear dependence of the dipole with energy is expected to happen at the highest energies. As we have shown, the existence of continuously emitting identical sources does not change the linear dependence of the dipole amplitude with the energy. Although, in this case, the interpretation of the normalization factor is not direct.

In Appendix~\ref{app:correlation} we show that the linear dependence of the dipole amplitude with the energy will be affected by a magnetic field with a coherence length about the particle gyroradius. The stochastic component destroys anisotropies of the distribution function. This effect is energy-dependent, breaking the linearity of the dipole amplitude with the energy. In figure~\ref{fig:dipole_diff} it is presented the effect of the diffusion on the fit of figure~\ref{fig:dipole}.

The sources of the UHECR are unknown~\citep{alves2019open} therefore it is impossible to be sure if they are isotropically and/or homogeneously distributed. The so-called top-down models propose relics of the Big Bang as sources of UHECRs, therefore isotropically and homogeneously distributed~\citep{PhysRevLett.79.4302}. However, top-down models are strongly disfavored by the data~(\cite{COLEMAN2023102819} and references therein). Data favors the hypothesis of astrophysical objects as sources of UHECRs, potentially active galactic nuclei, starburst galaxies, gamma-ray bursts, and/or other extreme astrophysical objects~\citep{aab2018indication}. These sources are distributed following the matter in the Universe, which is isotropic at cosmological scales (hundreds of megaparsecs). Therefore isotropic sources of UHECR is a hypothesis shared by the most accurate astrophysical models.

The homogeneity of the sources is not so easy to assure. For example, active galactic nuclei evolves with redshift as $(1+z)^{5}$ for $z<1.7$, while starburst galaxies evolves as $(1+z)^{3.4}$ for $z<1$~\citep{Gelmini_2012}. Besides the distribution of the sources, the propagation of the particles induces overdensities and horizons~\citep{Greisen1966,bib:zk,MLemoine_2009}. In the energy range between $1-10$~EeV, the horizon imposed by the energy losses is $\sim 10^3$ Mpc~\citep{lang2020revisiting,Ding_2021} depending on the UHECR type, and can be smaller if magnetic horizons are taken into account. The localized acceleration on astrophysical objects and the energy loss horizons disfavors the hypothesis of homogeneously distributed sources of UHECR. 

If the sources of UHECRs are isotropic and inhomogenous, the arrival direction of the particles on Earth can have a dipolar signal as shown by equations~\ref{eq:dipole} and~\ref{eq:dipole:weak:field}. In this case, the dipole measured by the Pierre Auger Observatory needed to be interpreted as a combination of inhomogeneous sources and magnetic fields.

%=======================================
\section{Quadrupole evolution} 
\label{sec:quadrupole}
Ignoring the octapole contribution, the quadrupole ($\ell = 2$) time evolution is given by
\begin{equation} 
\label{eq:quadrupole}
    \partial_t Q_{ij} - \frac{qc}{E}B_m ( \epsilon_{nmi} Q_{nj} + \epsilon_{nmj} Q_{in} ) = -2c \Big[ \frac{1}{2}(\partial_i D_j + \partial_j D_i) - \frac{1}{3} \mathbf{\nabla} \cdot \mathbf{D} \delta_{ij} \Big],
\end{equation}
where $\epsilon_{ijk}$ is the antisymetric Levi-Civita symbol, and $\delta_{ij}$ is the Kronecker delta.
As in the dipole case, the magnetic field rotates the quadrupole direction. The quadrupole source corresponds to the shear term of $\mathbf{\nabla} \mathbf{D}$, given by $ \frac{1}{2} (\partial_i D_j + \partial_j D_i) - \frac{1}{3} \mathbf{\nabla} \cdot \mathbf{D} \delta_{ij}$ \big. The average quadrupole amplitude is defined as $Q^2=\sum_{ij} Q_{ij}^2 / 9$. Since the quadrupole is traceless and symmetric, we get $\frac{1}{2} \partial_t Q^2  = - \frac{1}{9} c \sum_{ij} Q_{ij}(\partial_i D_j + \partial_j D_i)$. This means that the quadrupole amplitude can grow up only due to the shear of the dipole.

The stationary value of the quadrupole can be estimated as $Q/\Phi \sim \frac{1}{\alpha} (D/\Phi)^2$, where $\alpha = (L_1/L) (\mu_2/\sin \theta)$, $\mu_2$ takes into account the angle between $\mathbf{B}$ and $Q_{ij}$, and $L_1$ is the dipole scale of variation. This predicts that the quadrupole value goes with the square of the dipole value. 

The data measured by the Pierre Auger Observatory shows that $D/\Phi \ll 1$ for energies up to $40$~EeV, therefore our models can be used to predict a negligible value of the quadrupole in this energy range. Using equation~\ref{eq:dipole}, we have $BL\mu \sim 70$~nG Mpc, then $D \sim E/200 \text{ EeV}$, implying that the quadrupole has negligible value for energies below $E \sim 60-600$~EeV, for $\alpha = 0.1-10$. The current data does not show a significant quadrupole for energies below $40$~EeV~\citep{deAlmeida:20212Z,di20232022}.

\section{Conclusion}

We present a comprehensive solution of Liouville's equation for charged particles in magnetic fields. The solution is presented using a spherical harmonics expansion of the phase density function which facilitates the analysis of the arrival direction of particles on Earth.

The solutions presented in section~\ref{sec:harmonic_exp} and \ref{sec:monopole_dipole} are general and can be applied to a wide range of problems related to the isotropy of particles propagating in magnetic fields. Some previously known aspects of this issue are revisited and assured with rare clarity by the formalism used, such as the hierarchy in the evolution of poles and the rotating of the pole caused by the magnetic fields. Already in the general solution presented in these sections, previous misunderstandings are resolved concerning the time evolution of poles from initial spatially homogeneous and isotropic source distributions.

We showed that a spatially isotropic but inhomogeneous source distribution can generate higher order poles by interactions with the magnetic field. This was proven both for a transient state for generic magnetic fields as well under the weak field approximation without any assumption on the field structure. This particular calculation is enough to prove that the widely used hypothesis that an initially isotropic flux of particles cannot become anisotropic due to the action of a magnetic field is incomplete. Our calculations show that the correct affirmation needs at least one additional hypothesis: (i) \textit{an initially isotropic and \underline{homogeneous} flux of particles cannot become anisotropic due to the action of a magnetic field}; or (ii) \textit{an initially isotropic flux of particles cannot become anisotropic due to the action of a magnetic field \underline{for continuously-emitting sources} whose particle flux reached a steady-state}.

%Our calculations show that the correct hypothesis is \textit{an initially isotropic and \underline{homogeneous} flux of particles cannot become anisotropic due to the action of a magnetic field}.

The calculations developed here offers a framework to understand the linear dependence of the dipole amplitude with energy. We show the linear dependence can be understood as a match between the magnetic field intensity and the average distance of the sources. We also predict that the linear relation is a result of a composition that does not evolve or evolve slowly with energy. Our model predicts a dependence of the dipole amplitude with rigidity ($E/Z$) therefore for the highest energies, if the composition evolves, the linear dependence with energy will break. This result holds for sources with the same energy spectrum but is completely independent of the source time-injection.

The presence of a stochastic component on the magnetic field generates a diffusion process of the charged particles. This process tends to destroy anisotropies along the trajectory, the destruction being faster the larger is $\ell$ (figure \ref{fig:decay_length}). The destruction of the large-scale anisotropies by the stochastic components of extragalactic magnetic fields makes nearby sources specially relevant for the interpretation of both large- and small-scales anisotropies. Beside the results of Appendix~\ref{app:correlation} be approximated (since neglect transitions between successive multipoles and energy losses), they reinforce that the dipole and quadrupole must be generated inside the cosmological isotropy scale~$\sim100$~Mpc for energies above $\sim10^{19}$~eV.

The destruction of multipoles is also energy-dependent, with the decay length increasing with the energy. This effect breaks the linear dependence of the dipole amplitude with the energy, becoming relevant below a threshold were the diffusion dominates (figure~\ref{fig:dipole_diff}).

In section~\ref{sec:quadrupole}, we calculate the time evolution of the quadrupole. Under the stationary conditions, we show the quadrupole amplitude is related to the square of the dipole amplitude. Given the energy dependence of the dipole measure by the Pierre Auger Observatory, we predict the quadrupole amplitude will become dominant over the dipole for particles with energies between 60 and 600 EeV.

\section*{Acknowledgments}
CO and VdS acknowledge FAPESP Projects 2019/10151-2, 2020/15453-4, and all authors acknowledge FAPESP Project  2021/01089-1. The authors acknowledge the National Laboratory for Scientific Computing (LNCC/MCTI,  Brazil) for providing HPC resources for the SDumont supercomputer (http://sdumont.lncc.br). VdS acknowledges CNPq. This study was financed in part by the Coordenação de Aperfeiçoamento de Pessoal de Nível Superior - Brasil (CAPES) - Finance Code 001. Thanks to Darko Veberic for reading the manuscript and sending valuable comments.

\bibliography{main.bib}
%%%%%%%%%%Figure%%%%%%%%%%%%%
%------------------------------------
\newpage
%------------------------------------
\appendix
%------------------------------------
\section{Poles evolution in time according to Liouville's equation 
} 
\label{app:multipoles}
Starting from equation \ref{eq:liouville-projected},
\begin{equation}
  \sum_{\ell,m} \Bigg[ (\partial_t a_{\ell m}) Y_{\ell m} + c  (\hat{\mathbf{p}} \cdot \mathbf{\nabla} a_{\ell m}) Y_{\ell m} - i\frac{qc}{E} a_{\ell m} \mathbf{B} \cdot \mathbf{L} Y_{\ell m} \Bigg] = 0,
\end{equation}
the evolution of just one particular $a_{\ell m}$ is calculating by projecting over $Y^*_{\ell m}$. It is done by applying $\int_{\Omega_p} d\Omega_p Y^*_{\ell m}$ over the full equation
\begin{equation}
  \int_{\Omega_p} d\Omega_p \sum_{\ell,m} \Bigg[ (\partial_t a_{\ell m}) Y_{\ell m} Y^*_{\ell m} + c  (\hat{\mathbf{p}} \cdot \mathbf{\nabla} a_{\ell m}) Y_{\ell m} Y^*_{\ell m} - i\frac{qc}{E} a_{\ell m} \mathbf{B} \cdot \Big( \mathbf{L} Y_{\ell m} \Big) Y^*_{\ell m} \Bigg] = 0.
\end{equation}
For clarity, we analyze each term of the equation using the identification
\begin{eqnarray*}
A & = & \int_{\Omega_p} d\Omega_p \sum_{\ell,m} \Bigg[ (\partial_t a_{\ell m}) Y_{\ell m} Y^*_{\ell m} \Bigg] \\
B & = & \int_{\Omega_p} d\Omega_p \sum_{\ell,m} \Bigg[ c  (\hat{\mathbf{p}} \cdot \mathbf{\nabla} a_{\ell m}) Y_{\ell m} Y^*_{\ell m} \Bigg] \\
C & = & \int_{\Omega_p} d\Omega_p \sum_{\ell,m} \Bigg[ - i\frac{qc}{E} a_{\ell m} \mathbf{B} \cdot \Big( \mathbf{L} Y_{\ell m} \Big) Y^*_{\ell m} \Bigg] 
\end{eqnarray*}
such as that $A+B+C = 0$.
%%%%%
\subsection{Term $A$}
The orthogonality of the $Y_{\ell m}$ can be directly applied over Term $A$, resulting in
\begin{equation}
    A = \int_{\Omega_v} d\Omega_v \sum_{\ell',m'} (\partial_t a_{\ell' m'}) Y_{\ell' m'} Y^*_{\ell m} =
    \sum_{\ell',m'} (\partial_t a_{\ell' m'}) \int_{\Omega_v} d\Omega_v  Y_{\ell' m'} Y^*_{\ell m} = \partial_t a_{\ell m}
\end{equation}
%%%%
\subsection{Term $B$}
Writing $\hat{\mathbf{p}} = (p_x,p_y,p_z)$ in terms of the spherical harmonics
\begin{eqnarray*}
    p_x &= &\sin(\theta_p) \cos(\phi_p) = -\sqrt{2\pi/3} (Y_{11} - Y_{1 -1}), \\
    p_y &= &\sin(\theta_p) \sin(\phi_p) = i \sqrt{2\pi/3} (Y_{11} + Y_{1 -1}), \\
    p_z &= &\cos(\theta_p) = \sqrt{4\pi/3} Y_{10},
\end{eqnarray*}  
and replacing we obtain
\begin{align*}
    B = c \sqrt{\frac{4\pi}{3}} \sum_{\ell',m'} \int_{\Omega_p} d\Omega_p \bigg(& -\frac{(\partial_x - i\partial_y )}{\sqrt{2}}a_{\ell' m'} Y_{11} +\\
    &+ \frac{(\partial_x + i\partial_y)}{\sqrt{2}}a_{\ell' m'} Y_{1 -1} 
    + \partial_z a_{\ell' m'} Y_{1 0}
    \bigg) Y_{\ell' m'} Y_{\ell m}^*.
\end{align*}

Using the spherical harmonics property
\begin{equation} \label{eq:tripleYintegral}
    \int d\Omega_p Y_{l_1 m_1} Y_{l_2 m_2} Y_{l_3 m_3} = \sqrt{\frac{(2l_1 + 1)(2l_2 + 1)}{4\pi (2l_3 +1)}} (-1)^{-m_3} C^{l_1 0| l_2 0}_{l_3 0} C^{l_1 m_1| l_2 m_2}_{l_3 -m_3}
\end{equation}
where $C^{l_1 m_1| l_2 m_2}_{l_3 m_3} = \langle l_1 m_1 l_2 m_2 | l_3 m_3 \rangle$ are the Clebsch–Gordan coefficients. We get
\begin{align*}
    \int d\Omega_p Y_{1 k} Y_{\ell' m'} Y_{\ell m}^* &=
    (-1)^m \int d\Omega_p Y_{1 k} Y_{\ell' m'} Y_{\ell -m} =
    (-1)^{-m-m} \sqrt{\frac{3(2\ell' + 1)}{4\pi (2\ell +1)}} C^{1 0| \ell' 0}_{\ell 0} C^{1 k | \ell' m'}_{\ell m}\\
    & = \sqrt{\frac{3}{4\pi}} \sqrt{\frac{2\ell' + 1}{2\ell +1}} C^{1 0| \ell' 0}_{\ell 0} C^{1 k | \ell' m'}_{\ell m},
\end{align*}
which can be solved to
\begin{align*}
   B =  c \sum_{\ell',m'} \sqrt{\frac{2\ell'+1}{2\ell + 1}} C^{1 0| \ell' 0}_{\ell 0} & \Bigg[-\frac{(\partial_x - i\partial_y )}{\sqrt{2}} C^{1 1 | \ell' m'}_{\ell m} +\\
    &+ i\frac{(\partial_x + i\partial_y )}{\sqrt{2}} C^{1\ -1 | \ell' m'}_{\ell m} + C^{1 0 | \ell' m'}_{\ell m} \partial_z  \Bigg] a_{\ell' m'}.
\end{align*}
%%%%
\subsection{Term $C$}
Writing $\Vec{B} \cdot \Vec{L} = \frac{1}{2} \big( B_- L_+ + B_+ L_- \big) + B_z L_z$, where $B_\pm := B_x \pm i B_y$ and $L_\pm := L_x \pm i L_y$ and using the properties of the ladder operators over the $Y_{\ell m}$, we get
\begin{eqnarray*}
     C & = &\int_\Omega d\Omega \sum_{\ell',m'} a_{\ell' m'} \Vec{B}\cdot \Vec{L} Y_{\ell' m'} Y^{*}_{\ell m}  \\
     C & = &\sum_{\ell',m'} a_{\ell' m'} \delta_{\ell,\ell'} \bigg(\frac{B_-}{2}\sqrt{\ell'(\ell'+1) - m'(m'+1)} \delta_{m,m'+1} +  \\
     &  & \qquad + \frac{B_+}{2} \sqrt{\ell'(\ell'+1) - m'(m'-1)} \delta_{m,m'-1} + m'B_z \delta_{m,m'} \bigg) \\
     C & = &\bigg(\frac{B_-}{2}\sqrt{\ell(\ell+1) - m(m-1)} a_{\ell,m-1} + \\
     &  & \qquad + \frac{B_+}{2} \sqrt{\ell(\ell+1) - m(m+1)} a_{\ell,m+1} + m B_z a_{\ell,m} \bigg)
\end{eqnarray*}
%%%%
Combining the results from Terms A, B, and C, with the source term, we get equation~\ref{eq:liouville_expanded}.
%===========================================================
\section{Evolution of the amplitude of a pole without transition effects} 
\label{app:amplitude:evolution}
The amplitude of a pole is given by $\sum_{m=-\ell}^\ell |a_{\ell m}|^2$. Its evolution can be obtained multiplying equation~\ref{eq:liouville_expanded} by $a_{\ell m}^*$ and summing on $m$,
\begin{equation} 
\label{apB:1}
  \frac{1}{2} \sum_{m=-\ell}^\ell \partial_t |a_{\ell m}|^2 
  + \frac{qc}{E} \sum_{m=-\ell}^{\ell} \sum_{m'=-\ell}^{\ell} a_{\ell m}^* \mathrsfso{B}_{m}^{m'} a_{\ell m'} =
  - c \sum_{m=-\ell}^\ell \sum_{\ell' m'} a_{\ell m}^* T_{\ell m}^{\ell' m'} a_{\ell' m'}
\end{equation}

To analyze the effect of the magnetic field we cancel the transition tensor ($T_{\ell m}^{\ell' m'} = 0)$ in equation~\ref{eq:liouville_expanded} and focus on the magnetic field $\mathrsfso{B}$ term. This corresponds to an irrealistic condition in which all poles are identically null in all points of the space at any time. Physically speaking, the condition corresponds to the case in which no sources are considered.

The magnetic field term can be written
\begin{align*}
\label{eq:amp:b}
    &\sum_{m=-\ell}^{\ell} \sum_{m'=-\ell}^{\ell} a_{\ell m}^* \mathrsfso{B}_{m}^{m'} a_{\ell, m'} =\\
    &-i\sum_{m=-\ell}^{\ell} a_{\ell m}^* \bigg(\frac{B_-}{2}\sqrt{\ell(\ell+1) - m(m-1)} a_{\ell,m-1} +
     \frac{B_+}{2} \sqrt{\ell(\ell+1) - m(m+1)} a_{\ell,m+1} + m B_z a_{\ell,m} \bigg) =\\
     &\frac{1}{2i}\sum_{m=-\ell}^{\ell} \Bigg[ B_x \bigg( \sqrt{\ell(\ell+1) -m(m+1)}a_{\ell m}^* a_{\ell m+1} + \sqrt{\ell(\ell+1) -m(m-1)}a_{\ell m}^* a_{\ell m-1} \bigg) \\
     &+ B_y \bigg( \sqrt{\ell(\ell+1) -m(m+1)}a_{\ell m}^* a_{\ell m+1} - \sqrt{\ell(\ell+1) -m(m-1)}a_{\ell m}^* a_{\ell m-1} \bigg)  + 2m B_z |a_{\ell m}|^2 \Bigg]
\end{align*}

The magnetic field components $B_x (\mathbf{r})$, $B_y(\mathbf{r})$, and $B_z(\mathbf{r})$ are linearly independent. Using $a_{\ell m} \equiv b_m$ and $a_{\ell m}^* = (-1)^m a_{\ell -m}$, we can write for $B_z$ 

\begin{equation}
    \sum_{m=-\ell}^\ell 2m |b_m|^2 =
    \sum_{m=-\ell}^{-1} 2m |b_m|^2 + \sum_{m=1}^{\ell} 2m |b_m|^2 \\
    = \sum_{m=1}^{\ell} 2(m - m) |b_m|^2 = 0,
\end{equation}
since
\begin{equation*}
    |b_m|^2 = b_m b_m^* = (-1)^m b_{-m} (-1)^m b_{-m}^* = b_{-m} b_{-m}^* = |b_{-m}|^2.
\end{equation*}

The terms of $B_x$ and $B_y$ in equation~\label{eq:amp:b}
are different only by the minus signal on the $a_{\ell -m}$ term. They are shown together in the same equation with the $\pm$ sign representing the $+ \mapsto B_x$ and $- \mapsto B_y$
\begin{align*}
    &\sum_{m=-\ell}^{\ell} \Bigg[ \sqrt{\ell(\ell+1) -m(m+1)}b_{m}^* b_{m+1} \pm \sqrt{\ell(\ell+1) -m(m-1)}b_{m}^* b_{m-1} \Bigg] \\
    &= \sum_{m=-\ell}^{\ell} (-1)^m b_{-m} \Bigg[ \sqrt{\ell(\ell+1) -m(m+1)}b_{m+1} \pm \sqrt{\ell(\ell+1) -m(m-1)}b_{m-1} \Bigg] .
\end{align*}
Consider the terms $m=k$, $m=-(k-1)$, and $m=-(k+1)$. The sum of these terms is
\begin{align*}
    &(-1)^k \Bigg[ \underbrace{\sqrt{\ell(\ell+1) - k(k+1)} b_{-k} b_{k+1} \pm \sqrt{\ell(\ell+1) - k(k-1)} b_{-k}b_{k-1}}_{m=k}\\
    &\underbrace{- \sqrt{\ell(\ell+1) - (k-1)(k-2)} b_{k-1}b_{2-k} \mp \sqrt{\ell(\ell+1) - k(k-1)} b_{k-1} b_{-k} }_{m=-(k-1)}\\
    &\underbrace{-\sqrt{\ell(\ell+1) - k(k+1)} b_{k+1}b_{-k} \mp \sqrt{\ell(\ell+1) -(k+1)(k+2)} b_{k+1}b_{-(k+2)} }_{m=-(k+1)} \Bigg] \\
    &= (-1)^k \Bigg[ \underbrace{- \sqrt{\ell(\ell+1) - (k-1)(k-2)} b_{k-1}b_{2-k}}_{m=-(k-1)}  
    \underbrace{\mp \sqrt{\ell(\ell+1) -(k+1)(k+2)} b_{k+1}b_{-(k+2)} }_{m=-(k+1)} \Bigg]
\end{align*}
Note that all terms coming from $m=k$ are canceled. As this is valid for all $k$, it is only necessary to analyze the border terms, $m=\ell$, and $m=-\ell$. As $m=-(\ell+1)$ does not exist, $m=\ell$ only sum with $m=1-\ell$,
\begin{align*}
    &\underbrace{(-1)^\ell b_{-\ell} \Bigg[ \sqrt{\ell(\ell+1)- \ell(\ell+1)} b_{\ell+1} \pm \sqrt{\ell(\ell+1) - \ell(\ell-1)} b_{\ell-1} }_{m=\ell}\\
    &+ \underbrace{(-1)^{\ell+1} b_{\ell-1} \Bigg[ \sqrt{\ell(\ell+1) -(\ell-1)(\ell-2)}b_{2-\ell} \pm \sqrt{\ell(\ell+1) -\ell(\ell-1)}b_{-\ell} }_{m=1-\ell} \Bigg]\\
    &= \underbrace{(-1)^{\ell+1}b_{\ell-1} \sqrt{\ell(\ell+1) - (\ell-1)(\ell-2) b_{2-\ell}} }_{m=1-\ell}.
\end{align*}
The remaining term cancels out as before since it comes from $m=1-\ell$. A similar procedure to $m=-\ell$ leads to the same conclusion. As all terms in the sum cancel, the terms that multiply $B_x$ and $B_y$ are identically zero.

The angular power spectrum coefficients are defined as
\begin{equation}
    C_\ell = \frac{1}{2\ell + 1} \sum_{m=-\ell}^\ell |a_{\ell m}|^2.
\end{equation}
The evolution of $C_\ell$ can be taken from equation~\ref{apB:1}. According to the result presented here, the magnetic term is null, and then
\begin{equation} 
  \partial_t C_\ell =
  - \frac{2c}{2\ell + 1} \sum_{m=-\ell}^\ell \sum_{\ell' m'} a_{\ell m}^* T_{\ell m}^{\ell' m'} a_{\ell' m'}
\end{equation}

The contribution of $B_x$, $B_y$, and $B_z$ are all null. Therefore, under the irrealist hypothesis of $T_{\ell m}^{\ell' m'} = 0$ (no sources) all $C_\ell$ coefficients are invariant under the action of a magnetic field leading to a conservation of the amplitude of one pole under the action of a magnetic field. As discussed in section~\ref{sec:harmonic_exp} the magnetic field can cause only spatial rotations of the poles (i.e. rotation of the dipole direction), corresponding to transitions $m \mapsto m, m \pm 1$ coupled by the magnetic tensor.

%==================================================================
\section{Evolution of the monopole and dipole using the perturbative method}
\label{app:weak-field}

The evolution of the leading order and first perturbative order of the monopole and dipole are ruled by (equations \ref{eq:order_zero} and \ref{eq:order_one})
\begin{align}
    &\partial_t \Phi^{(0)} = - c \nabla \cdot \mathbf{D}^{(0)}, \label{eq:order_zero_m}\\
    &\partial_t \mathbf{D}^{(0)} = - \frac{c}{3} \mathbf{\nabla} \Phi^{(0)}, \label{eq:order_zero_d}\\
    &\partial_t \Phi^{(1)} = - c \nabla \cdot \mathbf{D}^{(1)}, \label{eq:order_one_m}\\
    &\partial_t \mathbf{D}^{(1)} = - \frac{c}{3} \mathbf{\nabla} \Phi^{(1)} - \frac{qc}{E} \mathbf{B} \times \mathbf{D}^{(0)} \label{eq:order_one_d}.
\end{align}

The $\Phi^{(0)}$ can be decoupled from $\mathbf{D}^{(0)}$ taking the temporal derivative of equation \ref{eq:order_zero_m} and substituting \ref{eq:order_zero_d}. A similar procedure decouples $\Phi^{(1)}$ from $\mathbf{D}^{(1)}$, resulting in
\begin{align}
    &\Big(\nabla^2 - \frac{3}{c^2} \partial_t^2 \Big) \Phi^{(0)} = 0, \label{eq:monopole0_weak2} \\
    &\mathbf{D}^{(0)} = \mathbf{D}^{(0)}(t_0) - \frac{c}{3} \int d\Tilde{t}
    \mathbf{\nabla} \Phi^{(0)}(\Tilde{t}), \label{eq:dipole0_weak2} \\
    &\Big(\nabla^2 - \frac{3}{c^2} \partial_t^2 \Big) \Phi^{(1)} = \frac{3q}{E} \mathbf{\nabla} \cdot (\mathbf{D}^{(0)} \times \mathbf{B}), \label{eq:monopole1_weak2} \\
    &\mathbf{D}^{(1)} = - \frac{c}{3} \int d\Tilde{t} \mathbf{\nabla} \Phi^{(1)} (\Tilde{t})  -\frac{qc}{E} \mathbf{B} \times \int d\Tilde{t} \mathbf{D}^{(0)}(\Tilde{t}). \label{eq:dipole1_weak2}
\end{align}

The leading order monopole $\Phi^{(0)}$ satisfies a homogeneous wave equation. We are concerned with the case in which the initial condition consists of a pure monopole ($\mathbf{D}^{(0)}(t_0) = 0$) with spherical symmetry. This monopole corresponds to a spherically symmetric wave propagating from infinity to an observer at the origin. In this case, $\Phi^{(0)} (t,r) = r^{-1}F(kr+\omega t)$, where $\omega^2 = k^2 \frac{c^2}{3}$, and $F$ is dependent on initial conditions. Substituting this expression on equation~\ref{eq:dipole0_weak2}
\begin{align*}
    \mathbf{D}^{(0)} (\mathbf{r},t) &= - \frac{c}{3} \hat{\mathbf{r}} \int_0^t d\Tilde{t} \partial_r \Phi^{(0)}(r,t)\\
    &= - \frac{c}{3} \hat{\mathbf{r}} \int_0^t d\Tilde{t} \Bigg[ \frac{\partial_r F(kr+\omega\Tilde{t})}{r} - \frac{F(kr+\omega \Tilde{t})}{r^2} \Bigg]\\
    &= - \frac{c}{3} \hat{\mathbf{r}} \Bigg[ \frac{k}{\omega r} \int_{kr}^{kr+\omega t} d\xi \partial_\xi F(\xi)  - \int_0^t d\Tilde{t} \frac{F(kr+\omega \Tilde{t})}{r^2} \Bigg]\\
    &= - \frac{c}{3} \hat{\mathbf{r}} \Bigg[ \frac{1}{r} \frac{\sqrt{3}}{c} \Big( F(kr+\omega t) - F(kr) \Big) - \int_0^t d\Tilde{t} \frac{F(kr+\omega \Tilde{t})}{r^2} \Bigg]\\
    &= - \frac{c}{3} \hat{\mathbf{r}} \Bigg[ \frac{\sqrt{3}}{c} \Big( \Phi^{(0)}(r,t) - \Phi^{(0)}(r,0) \Big) - \frac{1}{r} \int_0^t d\Tilde{t} \Phi^{(0)}(r,\Tilde{t}) \Bigg]
\end{align*}

Therefore, 
\begin{equation}
    \mathbf{D}^{(0)} (\mathbf{r},t) = \frac{g(r,t)}{r^2} \hat{\mathbf{r}} = \frac{c}{3} \frac{1}{r^2} \Bigg[ \int_0^t d\tau F(r,\tau) - \frac{\sqrt{3}}{c} r \Big( F(r,t) - F(r,0) \Big) \Bigg] \hat{\mathbf{r}} \label{eq:dip0_result}
\end{equation}

The first order perturbation of the monopole, $\Phi^{(1)}$, will be given by the source term of the inhomogeneous wave equation~\ref{eq:monopole1_weak2} (since it is a perturbation, the homogeneous solution is identified to $\Phi^{(0)}$). The solution can be written as~\citep{jackson1999classical}
\begin{align}
    \Phi^{(1)\pm}(\mathbf{r},t) = - \frac{1}{4\pi} \frac{3q}{E} \int \int d^3\mathbf{r'} dt' \delta \Big( t' - t \pm |\mathbf{r}-\mathbf{r'}|/c \Big) \frac{\nabla' \cdot \Big( \mathbf{D}^{(0)} (\mathbf{r'},t') \times \mathbf{B}(\mathbf{r'}) \Big)}{|\mathbf{r}-\mathbf{r'}|},
\end{align}
where $\nabla'$ represents the derivatives with respect to $\mathbf{r'}$. Note that $\Phi^{(1)}$ is only dependent on $|\mathbf{r}|$, since  any angular dependence is integrated on $\mathbf{r'}$. Using this fact in equation~\ref{eq:dipole1_weak2} together with equation~\ref{eq:dip0_result} results in the first order perturbation for the dipole 
\begin{align}
    \mathbf{D}^{(1)}(\mathbf{r},t) = -\frac{c}{3} \int_0^t d\tau \partial_r \Phi^{(1)}(r,\tau) \hat{\mathbf{r}} + \frac{qc}{E} \frac{\hat{\mathbf{r}} \times \mathbf{B}(\mathbf{r})}{r^2} \int \; g(r,\tau) d\tau .
\end{align}

%==================================================================
\section{Magnetic fields with coherence lengths}
\label{app:correlation}

Astrophysical magnetic fields can be approximated by a regular and a stochastic component, $\mathbf{B} = \mathbf{B_0} + \delta \mathbf{B}$. In this case, the distribution function is also split into an average and a fluctuating part, $f = \Bar{f} + \delta f$~\cite{PhysRevLett.112.021101,Battaner_2015,AHLERS2017184}. Equation \ref{eq:liouville} becomes
\begin{equation}
    \partial_t \Bar{f} + c \mathbf{\hat{p}} \cdot \nabla \Bar{f} - i \frac{qc}{E} (\mathbf{B_0} + \delta \mathbf{B}) \cdot \mathbf{L} \Bar{f} = \partial_t \delta f + c \mathbf{\hat{p}} \cdot \nabla \delta f - i \frac{qc}{E} (\mathbf{B_0} + \delta \mathbf{B}) \cdot \mathbf{L} \delta f \; .
\end{equation}
For simplicity, in the following discussion, we ignore the source term. Taking the ensemble average, the evolution of $\Bar{f}$ can be written as~\cite{AHLERS2017184}
\begin{equation} \label{eq:liouville_diffusion0}
    \partial_t \Bar{f} + c \mathbf{\hat{p}} \cdot \nabla \Bar{f} - i \frac{qc}{E} \mathbf{B_0} \cdot \mathbf{L} \Bar{f} = - i \frac{qc}{E} \langle \delta \mathbf{B} \cdot \mathbf{L} \delta f \rangle \; .
\end{equation}

In the BGK (Bhatnagar, Gross \& Krook) approximation~\cite{bhatnagar1954model}, the right-hand side term is treated as an effective viscous term, that can be written as~\cite{AHLERS2017184} $i \frac{qc}{E} \langle \delta \mathbf{B} \cdot \mathbf{L} \delta f \rangle \approx \nu \frac{\mathbf{L}^2}{2} \Bar{f}$, where $\nu$ is a relaxation rate that destroys anisotropies. We approximate the relaxation rate $\nu$ to the scattering length $\lambda_s \approx c/\nu$, with~\cite{lang2020_dipole}
\begin{align}
    \lambda_s (E) \approx \kappa l_c
    \begin{cases}%
        (R_L/l_c)^{1/3} \; R_L < l_c\\%
        (R_L/l_c)^{2} \; R_L \leq l_c
    \end{cases}
\end{align}
where $R_L = Z^{-1}(E/\text{EeV})(\text{nG}/B_0)$~Mpc, $\kappa = B_0^2 / \delta B^2$, $l_c$ is the coherence length of the magnetic field, and we are considering an isotropic diffusion.

Written
\begin{equation}
    \Bar{f}(\mathbf{x},\mathbf{p},t) = \sum_{\ell,m} \Bar{a}_{\ell m}(\mathbf{x},|\mathbf{p}|,t) Y_{\ell m}(\theta_p, \phi_p) \;,
\end{equation}
and following the same procedure of section~\ref{sec:harmonic_exp}, we obtain
\begin{equation}
  \label{eq:liouville+diffusion}
  \partial_t \Bar{a}_{\ell,m} = - c \sum_{\ell' m'}T_{\ell m}^{\ell' m'} \Bar{a}_{\ell' m'} - \frac{qc}{E} \sum_{m'} \mathrsfso{B}_{m}^{m'} \Bar{a}_{\ell, m'} -\frac{c}{2 \lambda_s (E)} \ell (\ell +1) \Bar{a}_{\ell,m},
\end{equation}
were we apply $\mathbf{L}^2 Y_{\ell m} = \ell(\ell+1) Y_{\ell m}$, and with the regular magnetic field $\mathbf{B}_0$ constituting the tensor $\mathrsfso{B}$.

\subsection{Multipoles destruction}
\begin{figure}
  \centering
  \includegraphics[width=0.5\columnwidth]{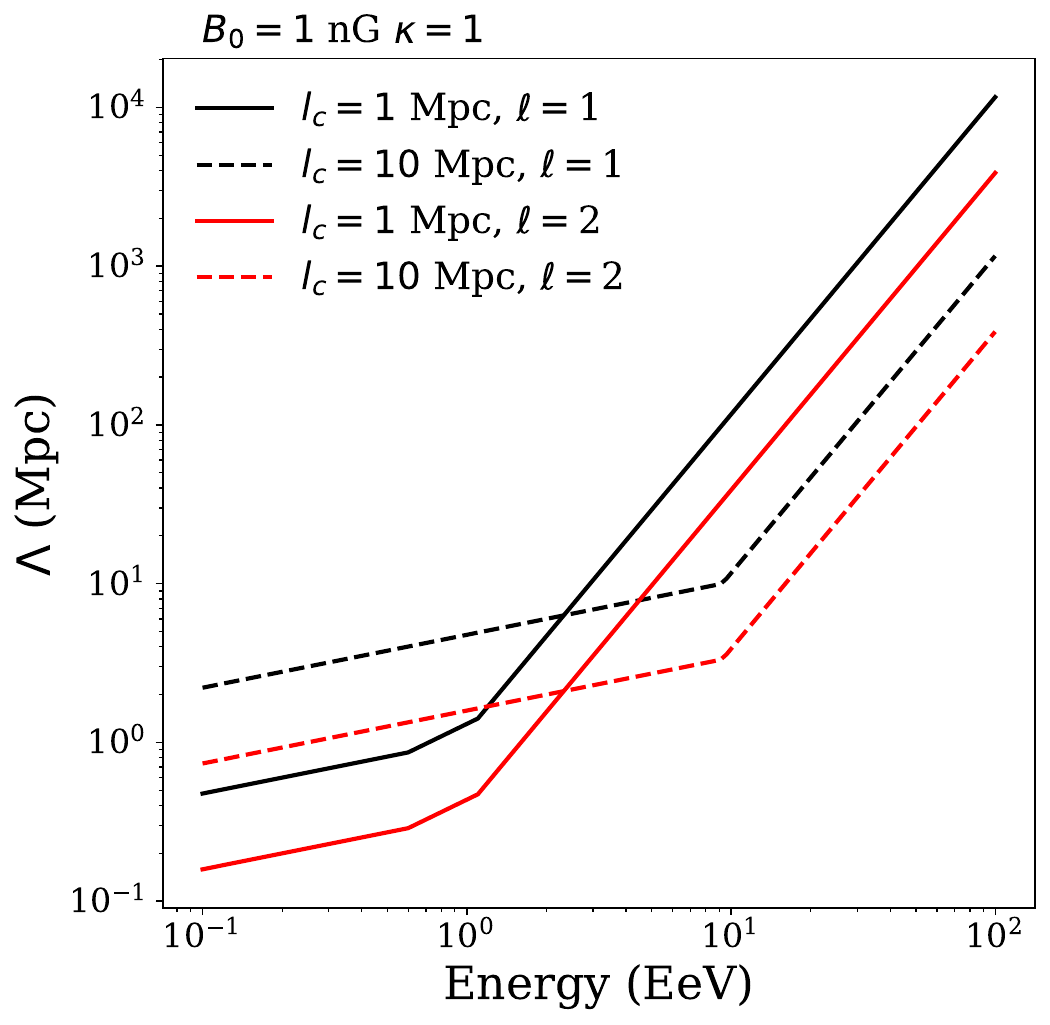}
  %161 x 297 mm = 297 words
  \caption{Decay length $\Lambda=2 \lambda_d / \ell(\ell+1)$ for $B_0 = 1$~nG and $l_c = 1, 10$~Mpc.}
  \label{fig:decay_length}
\end{figure}

To understand the effect of the BGK approximation, we consider that $T_{\ell m}^{\ell' m'} \Bar{a}_{\ell' m'} = 0$ and $\mathrsfso{B}_{m}^{m'} \Bar{a}_{\ell, m'} = 0$. Equation \ref{eq:liouville+diffusion} can be easily solved, leading to
\begin{equation}
    \Bar{a}_{\ell m}(t) = \Bar{a}_{\ell m}(t_0) e^{- \frac{\ell (\ell +1)}{2 \lambda_s (E)} c(t-t_0)}.
\end{equation}
The effect of the correlation term is to destroy anisotropies. Figure~\ref{fig:decay_length} presents the decay length defined as $\Lambda = 2 \lambda_s (E) / \ell(\ell+1)$ to the cases $\ell=1$ and $\ell=2$. For small ($\ll \Lambda$) propagation distances, the results obtained in the main text remain valid. Higher poles (small-scale anisotropies) are destroyed more quickly than large-scale anisotropies. Particles of smaller energies are more subject to the friction effect since they perform more revolutions in the magnetic field.

\subsection{Monopole and dipole evolution}
The evolution of the mean monopole and dipole are given by
\begin{align}
    &\partial_t \Phi (\mathbf{x},t) = - c \mathbf{\nabla} \cdot \mathbf{D}(\mathbf{x},t) \label{eq:monopole_diff}\\
    &\partial_t \mathbf{D}(\mathbf{x},t) = -\frac{c}{3} \mathbf{\nabla} \Phi +  \frac{qc}{E} \mathbf{D}(\mathbf{x},t) \times \mathbf{B}(\Vec{x}) - \frac{c}{\lambda_s (E)} \mathbf{D}(\mathbf{x},t)
    \label{eq:dipole_diff},
\end{align}
where we neglect the quadrupole term as before.

\subsubsection{Stationary solution}
Imposing stationarity in equation~\ref{eq:dipole_diff}, with $\nabla \Phi \approx \Phi/L$, we get
\begin{equation}
    \frac{D}{\Phi} = \frac{1}{3L} \Bigg[ \bigg(\frac{B_0 \mu}{E/Z}\bigg)^2 + \lambda_s^{-2} \Bigg]^{-1/2}
\end{equation}

\begin{figure}
  \centering
  \includegraphics[width=0.5\columnwidth]{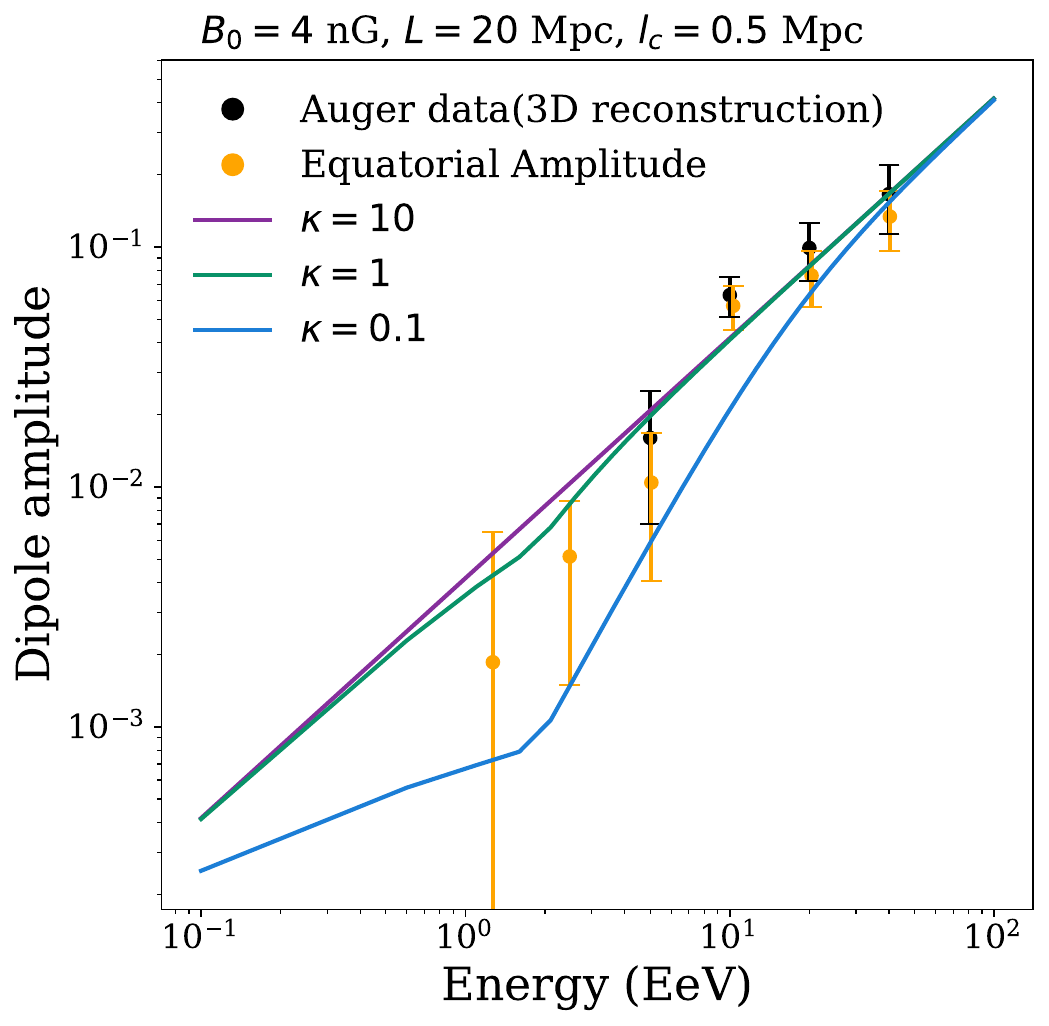}
  %161 x 297 mm = 297 words
  \caption{Effects of diffusion on the fit of the Auger dipole data~\cite{deAlmeida:20212Z}. For comparison, the data of the equatorial dipole amplitude~\cite{Aab_2020_raAnisotropies} are also shown.}
  \label{fig:dipole_diff}
\end{figure}

Figure \ref{fig:dipole_diff} shows the modification of the correlation term into the fit of the Auger data. As $R_L = E/ZB_0$ is the particles' gyroradius in the magnetic field, the effects coming from diffusion become important when $R_L \sim \lambda_s$.

\end{document}